\documentclass[amsmath, amssymb, aps, prd, showpacs, nofootinbib]{revtex4-2}

\usepackage{graphicx}
\usepackage{bm}
\usepackage[bookmarks=false]{hyperref}

\newcommand{\safeincludegraphics}[2][]{%
  \IfFileExists{#2}{\includegraphics[#1]{#2}}{\fbox{\texttt{\detokenize{#2}}}}%
}

\newcommand{\eqtag}[1]{\tag{#1}}

\newcommand{\ii}{\mathrm{i}}
\newcommand{\dd}{\mathrm{d}}

\begin{document}

\title{Feshbach--Villars Hamiltonian Approach to the Klein--Gordon Oscillator and Supercritical Step Scattering in Standard and Generalized Doubly Special Relativity}

\author{Abdelmalek Boumali}
\email{abdelmalek.boumali@univ-tebessa.dz}
\affiliation{Echahid Cheikh Larbi Tebessi University, Algeria}

\author{Nosratollah Jafari}
\affiliation{Fesenkov Astrophysical Institute, Almaty, Kazakhstan}

\author{Yassine Chargui}
\email{y.chergui@qu.edu.sa}
\affiliation{Department of Physics, College of Science, Qassim University, Buraydah 51452, Saudi Arabia}

\date{January 13, 2026}

\begin{abstract}
We develop a first-order Feshbach--Villars (FV) Hamiltonian framework for spin-0 relativistic quantum dynamics in the presence of Planck-scale kinematic deformations described within generalized doubly special relativity (G-DSR). Starting from a generic nonlinear momentum-space map, we derive the corresponding modified dispersion relation (MDR) at leading order in the Planck length \(l_p\) and construct a consistent FV linearization of the deformed Klein--Gordon operator. The resulting two-component Hamiltonian remains \(\sigma_3\)-pseudo-Hermitian at \(\mathcal{O}(l_p)\), which guarantees conservation of the FV charge and current and provides a current-based definition of reflection and transmission in stationary scattering.

As applications, we study two benchmark settings in which the FV metric structure is essential: (i) the one-dimensional Klein--Gordon oscillator and (ii) scattering from electrostatic step and barrier potentials. For the oscillator, we obtain controlled \(\mathcal{O}(l_p)\) branch-resolved spectral shifts and show how kinetic versus mass-shell deformations reshape the level spacing and the high-energy spectral compression. For step and barrier scattering, we compute reflection and transmission coefficients directly from the pseudo-Hermitian FV current and quantify the deformation-induced shift of the supercritical (pair-production) threshold. A comparative analysis of the Amelino--Camelia and Magueijo--Smolin realizations indicates that MS-type deformations generally delay the onset of the supercritical regime and reduce the magnitude of the negative transmitted flux within the validity domain \(l_p E \ll 1\).
\end{abstract}



\maketitle

\section{Introduction}
Relativistic quantum dynamics of spinless particles is governed by the Klein--Gordon (KG) equation, a second-order field equation that admits both positive- and negative-frequency solutions. While covariant and compact, its second-order structure obscures several aspects that are otherwise transparent in first-order formalisms: the construction of a conserved probability-like quantity, the identification of particle and antiparticle sectors, and the direct implementation of scattering boundary conditions.
A standard remedy is the two-component Feshbach--Villars (FV) representation, which rewrites the KG equation as a Schr\"odinger-like first-order system with a generally non-Hermitian Hamiltonian possessing a well-defined pseudo-Hermiticity property \cite{FV,Mostafazadeh2002,GreinerRQM}. This property induces an indefinite metric that yields a conserved charge density and a conserved current, and thereby provides a natural framework for scattering and spectral problems.

Historically, the FV construction can be viewed as the culmination of the single-particle interpretation program for the Klein--Gordon field that started with the charge--current reinterpretation of Pauli and Weisskopf \cite{PW34}. The first-order FV Hamiltonian \cite{FV} then enabled the systematic use of scattering boundary conditions and spectral techniques for scalar relativistic systems, and it was already employed early on in relativistic scattering theory \cite{Fuda1980}.

In recent years, the FV representation has enjoyed renewed attention in both flat and curved (or topologically nontrivial) backgrounds, where the positive-definite FV density is particularly advantageous. Representative flat-space applications include one-dimensional scattering analyses formulated directly in the FV framework \cite{Chargui2022}. On the curved/defect side, FV-based treatments have been developed for rotating or non-inertial cosmic-string geometries and related topological defects \cite{BouzenadaBoumali2023,NPBKK2023,GarahBoumali2025}.

In parallel, Planck-scale-motivated kinematics---often formulated in terms of doubly special relativity (DSR)---introduces an observer-independent high-energy scale $\kappa_{P}$ (typically associated with the Planck energy) in addition to the invariant speed of light \cite{AC2001,AC,MS,ACopen,KowalskiGlikman2005,ACphen2003}. Many DSR realizations can be described through non-linear maps between auxiliary (``linear'') and physical energy-momentum variables, leading to modified dispersion relations (MDRs) and deformed Hamiltonians. In quantum-mechanical applications, the key technical challenges are (i) to derive the MDR consistently from the chosen map and (ii) to linearize the deformed KG operator in a manner that preserves the physical conservation laws.

Beyond the early phenomenological analyses of modified dispersion relations in DSR \cite{ACphen2003}, several recent developments emphasize geometric formulations (curved momentum space, cotangent-bundle or ``rainbow'' geometries) and applications beyond strictly flat backgrounds \cite{CarmonaRelancio2021,Relancio2022}. These ideas have been explored, for example, in momentum-dependent black-hole geometries \cite{RelancioLiberati2022,Tao2022Rainbow}, in refined transformation schemes consistent with vanishing leading-order time delays \cite{Jafari2024}, and even in laboratory analog models \cite{Marino2025}.

The purpose of the present work is twofold. First, we provide a systematic FV formulation for a generic leading-order MDR in generalized DSR (G-DSR), including a proof that \(\sigma_3\)-pseudo-Hermiticity (and hence FV charge/current conservation) is preserved at \(\mathcal{O}(l_p)\) under a consistent symmetrized operator ordering. Second, we apply the resulting framework to benchmark problems where first-order structure and conserved currents are decisive: the one-dimensional KG oscillator and stationary scattering from electrostatic step and barrier potentials. In both cases, we compare the Amelino--Camelia and Magueijo--Smolin realizations and isolate the distinct mechanisms by which Planck-scale deformations modify spectral distributions and shift the onset of the supercritical (pair-production) regime.

To make the logic of the construction explicit, it is helpful to view the paper as an ``architecture'' with three layers: (i) a kinematic layer in which a momentum-space map fixes the MDR at leading order, (ii) a dynamical layer in which the deformed KG operator is linearized into an FV Hamiltonian that is pseudo-Hermitian with respect to the FV metric, and (iii) an observable layer in which spectra and scattering coefficients are extracted from the FV inner product and current. This separation clarifies where model-dependence enters (through the map parameters) and where consistency constraints must be imposed (through pseudo-Hermiticity and current conservation).

The paper is organized as follows. Section~\ref{sec:FV} reviews the FV representation, emphasizing pseudo-Hermiticity and the continuity equation. Section~\ref{sec:DSR} derives a generic leading-order MDR from a momentum-space map and constructs the corresponding FV Hamiltonian to $\mathcal{O}(l_p)$. Section~\ref{sec:KGO} treats the 1D KG oscillator and provides extended AC/MS spectral shifts. Section~\ref{sec:KleinParadox} studies the supercritical scattering for a step potential, derives $R$ and $T$ from the FV current, and analyzes the DSR-shifted threshold. Section~\ref{sec:discussion} discusses physical interpretation and limitations, and Section~\ref{sec:conclusion} summarizes the main conclusions.

\section[FV formalism]{Feshbach--Villars formalism and pseudo-Hermiticity}\label{sec:FV}
We briefly summarize the FV representation and its pseudo-Hermitian structure, which provides the conserved charge/current used throughout the paper \cite{FV,Mostafazadeh2002,GreinerRQM}.
We work in one spatial dimension for definiteness, but the FV construction is dimension-independent. Unless otherwise stated, we keep $\hbar$ and $c$ explicit to facilitate contact with nonrelativistic limits.

\subsection{From the Klein--Gordon equation to a first-order system}
Consider the free KG equation
\begin{equation}
\left(\frac{1}{c^2}\frac{\partial^2}{\partial t^2}-\frac{\partial^2}{\partial x^2}+\frac{m^2c^2}{\hbar^2}\right)\Phi(x,t)=0.
\label{eq:KGfree}
\end{equation}
The FV representation introduces a two-component wave function $\Psi=(\phi,\chi)^T$ through \cite{FV}
\begin{equation}
\phi=\frac{1}{2}\left(\Phi+\frac{\ii\hbar}{mc^2}\,\partial_t\Phi\right),\qquad
\chi=\frac{1}{2}\left(\Phi-\frac{\ii\hbar}{mc^2}\,\partial_t\Phi\right),
\label{eq:FVdef}
\end{equation}
which can be inverted as
\begin{equation}
\Phi=\phi+\chi,\qquad
\ii\hbar\,\partial_t\Phi=mc^2(\phi-\chi).
\label{eq:FVinverse}
\end{equation}
Using Eq.~(\ref{eq:KGfree}) to eliminate $\partial_t^2\Phi$ and Eq.~(\ref{eq:FVinverse}) to express time derivatives in terms of $\phi$ and $\chi$, one obtains a first-order equation
\begin{equation}
\ii\hbar\,\partial_t\Psi=\hat H_{\!\rm FV}\,\Psi,
\label{eq:FVsch}
\end{equation}
with FV Hamiltonian
\begin{equation}
\hat H_{\!\rm FV}=\left(\sigma_3+\ii\sigma_2\right)\frac{\hat p^{\,2}}{2m}+mc^2\sigma_3,
\qquad\hat p=-\ii\hbar\,\partial_x.
\label{eq:HFV}
\end{equation}
The matrix structure mixes $\phi$ and $\chi$, encoding positive- and negative-frequency sectors in a single two-component object.

\subsection{Pseudo-Hermiticity, conserved charge, and FV current}
The FV Hamiltonian is not Hermitian in the standard $L^2$ inner product. Instead, it obeys a $\sigma_3$-pseudo-Hermiticity condition
\begin{equation}
\hat H_{\!\rm FV}^{\dagger}=\sigma_3\,\hat H_{\!\rm FV}\,\sigma_3.
\label{eq:sigma3herm}
\end{equation}
This implies Hermiticity with respect to the indefinite inner product
\begin{equation}
\langle\Psi_1,\Psi_2\rangle_{\!\rm FV}=\int \dd x\,\Psi_1^{\dagger}(x)\,\sigma_3\,\Psi_2(x),
\label{eq:FVinner}
\end{equation}
and yields the conserved density
\begin{equation}
\rho(x,t)=\Psi^{\dagger}\sigma_3\Psi=|\phi|^2-|\chi|^2.
\label{eq:rho}
\end{equation}
To derive the continuity equation, multiply Eq.~(\ref{eq:FVsch}) on the left by $\Psi^{\dagger}\sigma_3$, subtract the Hermitian conjugate, and use Eq.~(\ref{eq:sigma3herm}). One finds
\begin{equation}
\partial_t\rho+\partial_x j=0,
\label{eq:cont}
\end{equation}
with FV current (in 1D)
\begin{equation}
 j(x,t)=\frac{\hbar}{2mi}\left[\Psi^{\dagger}(\sigma_3+\ii\sigma_2)\,\partial_x\Psi-\partial_x\Psi^{\dagger}(\sigma_3-\ii\sigma_2)\,\Psi\right].
\label{eq:current}
\end{equation}
In stationary scattering, Eq.~(\ref{eq:cont}) implies that the asymptotic currents are equal, leading to the flux-conservation identity $R+T=1$ when reflection $R$ and transmission $T$ are defined from the ratios of asymptotic currents.

\subsection{Couplings and operator ordering}
For external scalar potentials $V(x)$ implemented by the stationary substitution $E\to E-V(x)$ in the KG equation, the FV Hamiltonian inherits $V(x)$ as a diagonal term $V\,\mathbb{I}$ in the first-order system, while kinetic operators require attention to ordering when deformations introduce non-linear energy factors. In the present work, we consistently keep only leading $\mathcal{O}(l_p)$ terms and adopt symmetrization for mixed products such as $E\,\hat p^{\,2}$ to preserve pseudo-Hermiticity.

\section[Generalized DSR]{Generalized DSR and FV linearization}\label{sec:DSR}
Generalized DSR models are conveniently encoded by nonlinear momentum-space maps that preserve a relativity principle while introducing an invariant high-energy scale, leading to MDRs and deformed generators \cite{AC2001,AC,MS,ACopen,KowalskiGlikman2005,ACphen2003}.
\subsection{Momentum-space map and a generic leading-order MDR}
A convenient way to encode DSR kinematics is to introduce a non-linear map between auxiliary variables $(E,p)$ and physical variables $(P_0,P)$ that transform linearly under ordinary Lorentz transformations. To leading order in $l_p$ one may write the most general rotationally symmetric map consistent with $P_0\to E$ and $P\to p$ as $l_p\to 0$:
\begin{equation}
P_0\simeq E\left(1+\alpha_2 l_p E\right),\qquad
P\simeq p\left[1+(\alpha_3-\alpha_1) l_p E\right].
\label{eq:map}
\end{equation}
The constants $\alpha_1,\alpha_2,\alpha_3$ label the chosen realization. Imposing invariance of the auxiliary mass-shell,
\begin{equation}
P_0^2-P^2=m^2c^2, 
\label{eq:massshell}
\end{equation}
and substituting Eq.~(\ref{eq:map}) gives, to $\mathcal{O}(l_p)$,
\begin{align}
E^2-p^2c^2
-2\alpha_2 l_p E^3
+2(\alpha_3-\alpha_1) l_p E\,p^2c^2
&=m^2c^4.
\label{eq:MDR}
\end{align}
Equation~(\ref{eq:MDR}) is the generalized MDR used throughout. For later convenience we define $\Delta\alpha\equiv \alpha_3-\alpha_1$.

\subsection{Deformed KG operator and FV Hamiltonian to $\mathcal{O}(l_p)$}
To construct a quantum Hamiltonian, we promote $E\to \hat E\equiv \ii\hbar\,\partial_t$ and $p\to \hat p\equiv -\ii\hbar\,\partial_x$. With a scalar potential step $V(x)$, we use the stationary replacement $\hat E\to \hat E-V(x)$ in the MDR. The deformed KG operator then reads
\begin{align}
\Big[(\hat E-V)^2-\hat p^{\,2}c^2-m^2c^4\Big]\Phi
&=l_p\Big[2\alpha_2(\hat E-V)^3-2\Delta\alpha\,(\hat E-V)\hat p^{\,2}c^2\Big]\Phi,
\label{eq:deformedKG}
\end{align}
where equality holds up to $\mathcal{O}(l_p^2)$ terms.

The FV linearization proceeds by retaining the same two-component definition in Eq.~(\ref{eq:FVdef}) but replacing $\partial_t\Phi$ using the deformed equation order-by-order. An efficient approach is to write the first-order Hamiltonian as a deformation of $\hat H_{\!\rm FV}$,
\begin{equation}
\hat H_{\!\rm G\text{-}DSR}=\hat H_{\!\rm FV}+l_p\,\delta\hat H+\mathcal{O}(l_p^2),
\label{eq:Hpert}
\end{equation}
and determine $\delta\hat H$ by matching the squared operator $(\hat H_{\!\rm G\text{-}DSR})^2$ to the deformed KG operator. To first order, one finds
\begin{equation}
\delta\hat H = \alpha_2\,\hat E^2\,\sigma_3
-\Delta\alpha\,\frac{1}{2m}\,\Big\{\hat E,\,\hat p^{\,2}\Big\}\,(\sigma_3+\ii\sigma_2),
\label{eq:dH}
\end{equation}
where $\{A,B\}=AB+BA$ is an anticommutator and $\hat E$ is to be understood as the (stationary) energy eigenvalue in time-independent problems. In the free case, $\hat E\to E$ commutes with $\hat p$, and Eq.~(\ref{eq:dH}) reduces to the compact form
\begin{equation}
\hat H_{\!\rm G\text{-}DSR}=\hat H_{\!\rm FV}
+l_p\left[\alpha_2 E^2\sigma_3-\Delta\alpha\,\frac{E\,\hat p^{\,2}}{m}(\sigma_3+\ii\sigma_2)\right].
\label{eq:Hdsr}
\end{equation}

\subsection{Pseudo-Hermiticity and flux conservation in G-DSR}
A central consistency requirement is that the deformed Hamiltonian preserves a conserved FV charge at the working order. Using $\hat H_{\!\rm FV}^{\dagger}=\sigma_3\hat H_{\!\rm FV}\sigma_3$ and noting that $\sigma_3(\sigma_3+\ii\sigma_2)\sigma_3=(\sigma_3-\ii\sigma_2)$, it is straightforward to verify that the symmetrized deformation in Eq.~(\ref{eq:dH}) satisfies
\begin{equation}
\hat H_{\!\rm G\text{-}DSR}^{\dagger}=\sigma_3\,\hat H_{\!\rm G\text{-}DSR}\,\sigma_3+\mathcal{O}(l_p^2).
\label{eq:pseudo}
\end{equation}
Therefore the continuity equation and the FV current remain valid at $\mathcal{O}(l_p)$, guaranteeing flux conservation in scattering problems and the identity $R+T=1$ when $R$ and $T$ are defined via FV currents.

\subsection{AC and MS realizations as parameter choices}
The generalized MDR in Eq.~(\ref{eq:MDR}) contains as special cases commonly used DSR realizations. In the present leading-order parametrization, we identify:
\begin{itemize}
\item \textbf{AC-type:} $\alpha_2=0$ and $\Delta\alpha=1$ (dominant kinetic deformation).
\item \textbf{MS-type:} $\alpha_2=1$ and $\Delta\alpha=1$ (mass-shell shift and kinetic deformation of comparable order).
\end{itemize}
These assignments reproduce the qualitative features typically attributed to the AC and MS constructions: a purely kinetic leading correction in AC-type models and an additional energy-dependent ``mass-shell'' shift in MS-type models. The subsequent sections keep $\alpha_2$ and $\Delta\alpha$ symbolic to maintain generality, and then specialize to the above cases for extended comparisons.

% ----------------------------------------------------------------------
% IV. 1D KLEIN--GORDON OSCILLATOR IN G-DSR  (REVISED + CORRECTED)
% Natural units: \hbar=c=1.
% Changes applied:
%   (i) Keep operator ordering via A=x.
%   (ii) Remove any s=\pm 1 case-splitting.
%   (iii) Correct Eq. (21): \Lambda_n = m\omega\,2n.
%   (iv) Propagate this correction consistently through Eqs. (22)–(31).
%   (v) Remove all \boxed{...}.
% ----------------------------------------------------------------------
\section[Infinite-wall confinement]{Infinite-wall confinement in the FV formalism: SR, DSR, and G-DSR spectra}
\label{sec:infinitewell_FV}

The one-dimensional infinite square well provides a canonical testing ground for relativistic wave equations. In the present context it is particularly useful because the confinement fixes the spatial spectrum in a model-independent manner, allowing deformation-induced changes to be traced unambiguously to modifications of the energy--momentum relation \cite{GreinerRQM}. Throughout, we work in natural units $\hbar=c=1$.

\subsection{FV stationary problem in an infinite square well}
We consider an infinite square well of width $L$,
\[
V(x)=
\begin{cases}
0, & 0<x<L,\\
+\infty, & x\le 0 \ \text{or}\ x\ge L.
\end{cases}
\]
In the Feshbach--Villars (FV) representation, the two-component wave function $\Psi(x,t)$ satisfies a Schr\"odinger-type evolution equation,
\begin{equation}
i\frac{\partial}{\partial t}\Psi(x,t)=\hat H_{FV}\Psi(x,t),
\label{eq:FV_Schro}
\end{equation}
where the one-dimensional free FV Hamiltonian reads
\begin{equation}
\hat H_{FV}=(\sigma_3+i\sigma_2)\frac{\hat p^{\,2}}{2m}+m\sigma_3,
\qquad \hat p=-i\frac{d}{dx}.
\label{eq:FV_free}
\end{equation}
For stationary states $\Psi(x,t)=e^{-iEt}\Psi(x)$, Eq.~\eqref{eq:FV_Schro} reduces to the eigenvalue problem
\begin{equation}
\hat H_{FV}\Psi(x)=E\Psi(x).
\label{eq:FV_eig}
\end{equation}

The impenetrable walls enforce vanishing probability flux through $x=0$ and $x=L$. A convenient sufficient implementation, consistent with the FV inner product and ensuring self-adjointness on the confined domain, is to impose Dirichlet boundary conditions on the FV spinor,
\begin{equation}
\Psi(0)=0,\qquad \Psi(L)=0.
\label{eq:FV_bc}
\end{equation}
Inside the well ($0<x<L$), the dynamics are free and the spatial dependence reduces to the standard second-order form. The boundary conditions quantize the wave number as
\begin{equation}
k_n=\frac{n\pi}{L},\qquad n=1,2,3,\dots,
\label{eq:kn_FV}
\end{equation}
with eigenfunctions proportional to $\sin(k_n x)$ (up to FV spinor amplitudes fixed by the algebraic structure of $\hat H_{FV}$).

\subsection{Special-relativistic (SR) FV spectrum}
Squaring the FV eigenvalue equation (equivalently, reverting to the Klein--Gordon mass-shell condition) yields
\begin{equation}
E^2=m^2+k^2.
\label{eq:SR_disp}
\end{equation}
Thus, the confined SR energies are
\begin{equation}
E^{\rm (SR)}_{n,\pm}=\pm\sqrt{m^2+k_n^2}
=\pm\sqrt{m^2+\left(\frac{n\pi}{L}\right)^2}.
\label{eq:En_SR_FV}
\end{equation}
The two branches correspond to the particle/antiparticle sectors in the FV representation and are distinguished by the sign structure of the FV norm $\langle\Psi|\sigma_3|\Psi\rangle$.

\begin{figure}[t]
  \centering
  \safeincludegraphics[width=0.85\linewidth]{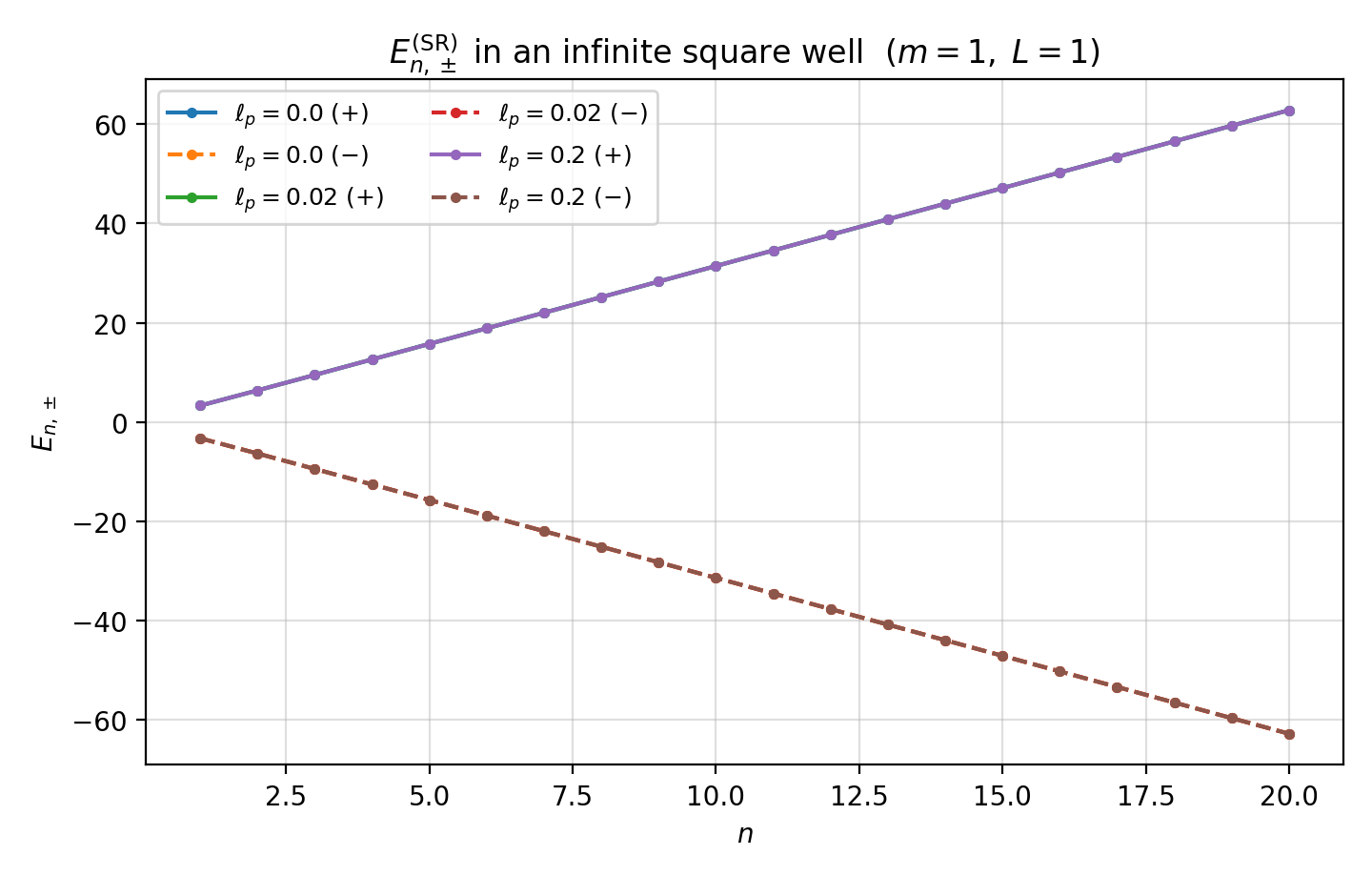}
  \caption{SR energy spectrum in an infinite square well for both FV branches $E^{(\mathrm{SR})}_{n,\pm}$ as a function of $n$, for $(m=1,\;L=1)$ and $\ell_p\in\{0,\;0.02,\;0.2\}$. The curves overlap because $\ell_p$ does not enter the SR dispersion relation.}
  \label{fig:SR_eq33_spectrum}
\end{figure}

\subsection{Standard DSR in FV: MS-type deformation of the mass shell}
To incorporate a standard DSR modification within the FV setting, we deform the energy variable entering the mass-shell relation while keeping the confinement-induced quantization \eqref{eq:kn_FV} unchanged. Adopting an MS-type nonlinear map,
\begin{equation}
E_{\rm DSR}=\frac{E}{1-l_pE},
\label{eq:MS_map_FV}
\end{equation}
one has $E_{\rm DSR}=E+l_pE^2+O(l_p^2)$. The stationary FV condition is then encoded in the deformed shell
\begin{equation}
E_{\rm DSR}^2=m^2+k_n^2.
\label{eq:DSR_shell_FV}
\end{equation}
Defining $\Omega_n\equiv\sqrt{m^2+k_n^2}$, Eq.~\eqref{eq:DSR_shell_FV} implies $E_{\rm DSR}=\pm\Omega_n$. Inverting \eqref{eq:MS_map_FV} yields
\begin{equation}
E^{\rm (DSR)}_{n,\pm}=\pm\frac{\Omega_n}{1+l_p\,\Omega_n},
\qquad
\Omega_n=\sqrt{m^2+\left(\frac{n\pi}{L}\right)^2}.
\label{eq:En_DSR_FV}
\end{equation}
Relative to SR, the deformation suppresses the magnitude of the energy levels and introduces an upper bound as $\Omega_n$ increases.

\begin{figure}[t]
  \centering
  \safeincludegraphics[width=0.85\linewidth]{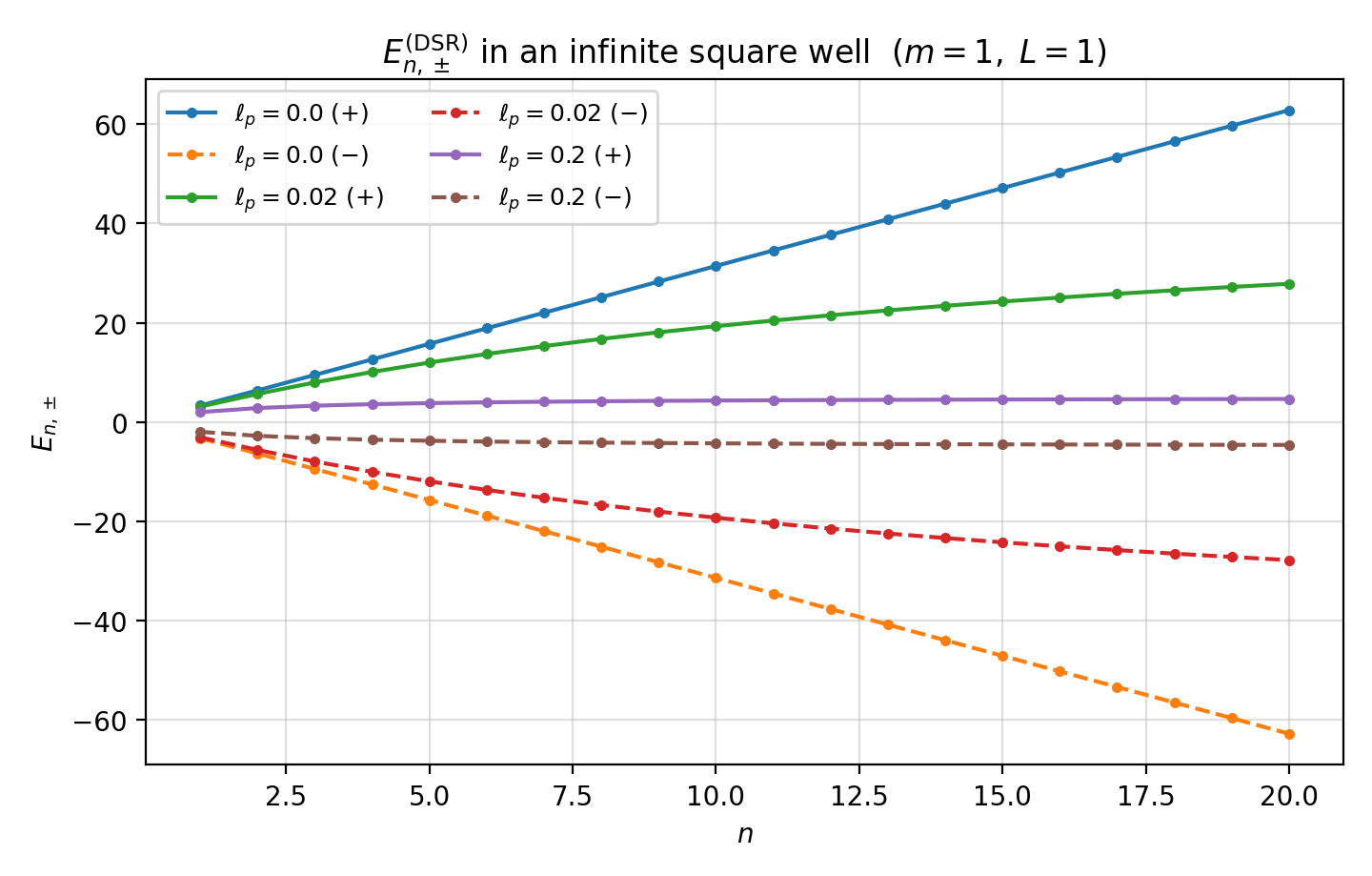}
  \caption{DSR energy spectrum in an infinite square well for both FV branches $E^{(\mathrm{DSR})}_{n,\pm}$ as a function of $n$, for $(m=1,\;L=1)$ and $\ell_p\in\{0,\;0.02,\;0.2\}$. Increasing $\ell_p$ suppresses the magnitude of the levels and induces a high-$n$ saturation.}
  \label{fig:DSR_eq34_spectrum}
\end{figure}

\subsection{Generalized DSR in FV: first-order flexible deformation}
For generalized DSR (G-DSR), we introduce a perturbative deformation compatible with the FV stationary structure,
\begin{equation}
E_{\rm eff}=E\left(1+\chi\,l_pE\right),
\label{eq:GDSR_map_FV}
\end{equation}
so that $E_{\rm eff}=E+\chi l_pE^2+O(l_p^2)$. The corresponding shell condition is
\begin{equation}
E_{\rm eff}^2=m^2+k_n^2=\Omega_n^2.
\label{eq:GDSR_shell_FV}
\end{equation}
Setting $E_{\rm eff}=\pm\Omega_n$ and solving \eqref{eq:GDSR_map_FV} for $E$ gives $\chi l_pE^2+E\mp\Omega_n=0$. Selecting the branch that recovers SR in the limit $l_p\to0$, and focusing on $\chi>0$, one may write
\begin{equation}
E^{\rm (GDSR)}_{n,+}
=
\frac{-1+\sqrt{1+4\chi l_p\,\Omega_n}}{2\chi l_p},
\qquad
E^{\rm (GDSR)}_{n,-}
=
-\frac{-1+\sqrt{1+4\chi l_p\,\Omega_n}}{2\chi l_p}.
\label{eq:En_GDSR_FV}
\end{equation}

\begin{figure}[t]
  \centering
  \safeincludegraphics[width=0.85\linewidth]{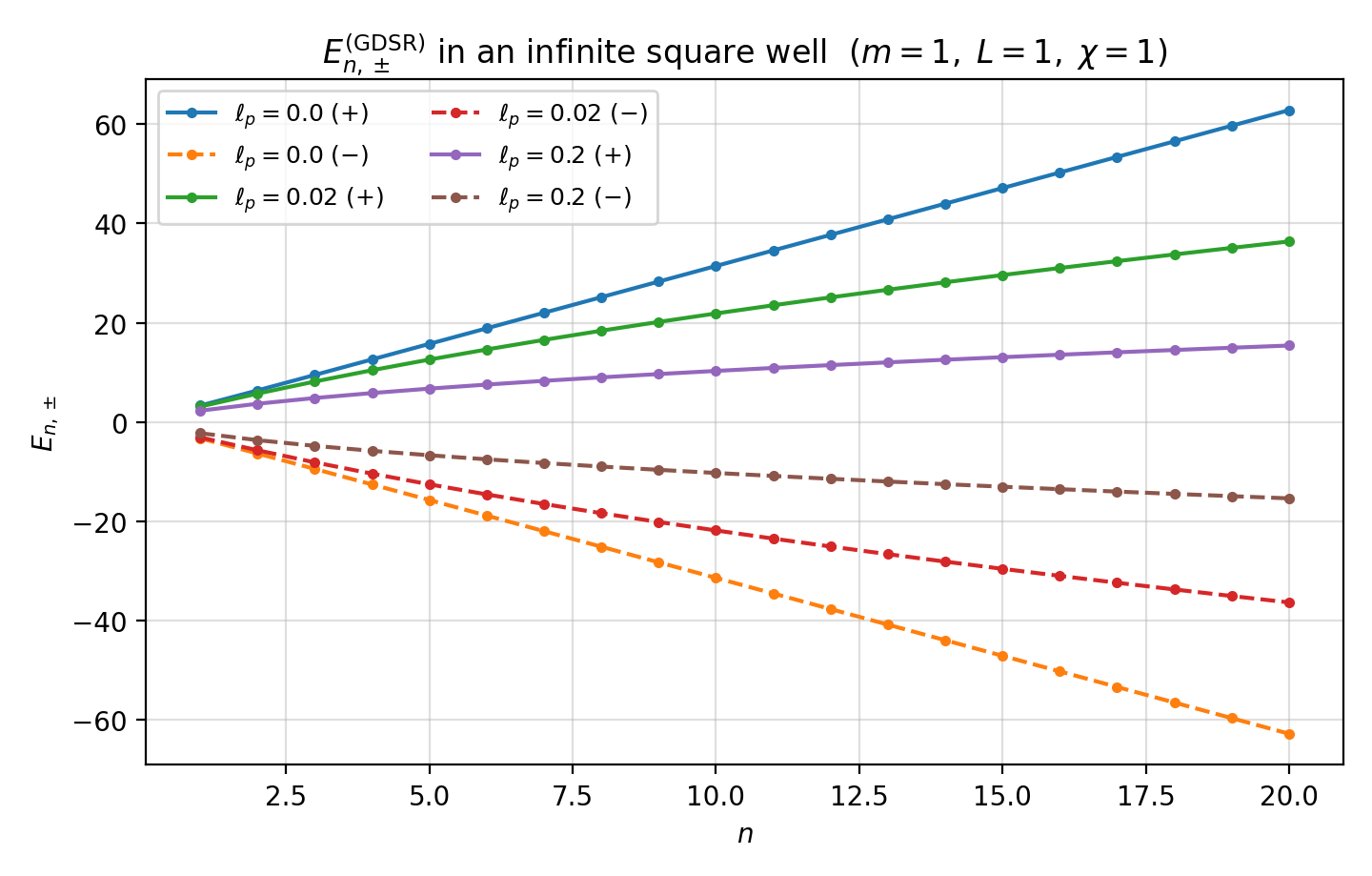}
  \caption{G-DSR energy spectrum in an infinite square well for both FV branches $E^{(\mathrm{GDSR})}_{n,\pm}$ as a function of $n$, for $(m=1,\;L=1,\;\chi=1)$ and $\ell_p\in\{0,\;0.02,\;0.2\}$. The deformation lowers the energies relative to SR while preserving an unbounded spectrum.}
  \label{fig:GDSR_eq35_spectrum}
\end{figure}

\subsection{Results and discussion}
The infinite-wall confinement fixes the wave-number sequence \eqref{eq:kn_FV} independently of the deformation model. Consequently, SR, DSR, and G-DSR share identical spatial quantization and differ only through the dispersion relation that links the quantized momentum $k_n$ to the stationary energy. Collecting the results,
\begin{align}
E^{\rm (SR)}_{n,\pm} &= \pm\Omega_n,\label{eq:RD_SR}\\
E^{\rm (DSR)}_{n,\pm} &= \pm\frac{\Omega_n}{1+l_p\Omega_n},\label{eq:RD_DSR}\\
E^{\rm (GDSR)}_{n,\pm} &= \pm\frac{-1+\sqrt{1+4\chi l_p\,\Omega_n}}{2\chi l_p},\label{eq:RD_GDSR}
\end{align}
with $\Omega_n=\sqrt{m^2+(n\pi/L)^2}$. In the FV formulation the $\pm$ branches represent particle/antiparticle sectors associated with the indefinite FV inner product, and the symmetry $E_{n,-}=-E_{n,+}$ is maintained by construction in all three models considered here.

\paragraph{Special-relativistic benchmark.}
In SR, $|E^{\rm (SR)}_{n,\pm}|$ increases monotonically with $n$ and is asymptotically linear for highly excited modes:
for $n\gg mL/\pi$, $\Omega_n\simeq k_n$ and hence $|E^{\rm (SR)}_{n,\pm}|\sim n\pi/L$. The corresponding level spacing approaches a constant, $\Delta E_n \equiv E_{n+1,+}-E_{n,+}\to \pi/L$, so the high-energy density of states in the box is essentially uniform in $n$. In the opposite regime $k_n\ll m$, the positive-energy branch admits the familiar nonrelativistic expansion
\begin{equation}
E^{\rm (SR)}_{n,+}=m+\frac{k_n^2}{2m}+O\!\left(\frac{k_n^4}{m^3}\right),
\end{equation}
showing explicitly how the FV description reduces to the Schr\"odinger spectrum up to the additive rest energy.

\paragraph{MS-type DSR: energy saturation and spectral compression.}
The MS-type deformation produces two robust qualitative effects visible in Fig.~\ref{fig:DSR_eq34_spectrum}: (i) suppression of $|E_{n,\pm}|$ relative to SR at fixed $n$, and (ii) saturation at high excitation. Indeed, from \eqref{eq:RD_DSR},
\begin{equation}
\lim_{n\to\infty}E^{\rm (DSR)}_{n,\pm}=\pm\frac{1}{l_p},
\end{equation}
so the spectrum is bounded in magnitude by $1/l_p$ despite the unbounded growth of $k_n$. This induces a progressive compression of high-$n$ levels: the spacing satisfies $\Delta E_n\to 0$ as $n\to\infty$, signalling an accumulation of eigenvalues near the deformation scale. Expanding at large $\Omega_n$ gives
\begin{equation}
E^{\rm (DSR)}_{n,\pm}=\pm\left(\frac{1}{l_p}-\frac{1}{l_p^2\Omega_n}+O(\Omega_n^{-2})\right),
\end{equation}
which quantifies the approach to the plateau. From a physical perspective, the nonlinear map effectively implements a maximal attainable energy in the stationary spectrum, so that increasingly short-wavelength modes are realized primarily through changes in the FV spinor structure rather than through further increases in $|E|$. In thermodynamic applications (e.g.\ partition sums built from the discrete spectrum), see, e.g.,\ \cite{BoumaliJafariEPJC2025,BoumaliJafariPLB2025}, this boundedness is expected to soften UV contributions relative to SR because high-$n$ states do not carry arbitrarily large energy, even though their count still grows with $n$.

\paragraph{G-DSR: unbounded spectrum with softened growth.}
By contrast, the perturbative G-DSR deformation in \eqref{eq:RD_GDSR} preserves an unbounded spectrum while reducing the growth rate relative to SR (Fig.~\ref{fig:GDSR_eq35_spectrum}). For $\chi>0$ and $\Omega_n\to\infty$ one finds
\begin{equation}
E^{\rm (GDSR)}_{n,+}\simeq \sqrt{\frac{\Omega_n}{\chi l_p}}-\frac{1}{2\chi l_p}
\qquad (n\to\infty),
\end{equation}
and hence $E_{n,+}$ grows sublinearly with $n$ (since $\Omega_n\sim n$ at large $n$, the leading behavior is $E_{n,+}\sim \sqrt{n}$). The level spacing therefore decreases as $n$ increases,
\begin{equation}
\Delta E_n \sim \frac{1}{\sqrt{n}},
\end{equation}
indicating spectral compression without the formation of a finite-energy plateau. In this sense, G-DSR interpolates between SR (linear growth and asymptotically constant spacing) and MS-type DSR (bounded growth and vanishing spacing), providing a flexible phenomenological framework in which the UV behavior can be softened without enforcing a strict maximum energy.

\paragraph{Weak-deformation and nonrelativistic limits.}
For $l_p\Omega_n\ll 1$ both deformed models reduce perturbatively to SR. From \eqref{eq:RD_DSR} one obtains
\begin{equation}
E^{\rm (DSR)}_{n,\pm}=\pm\Omega_n\left(1-l_p\Omega_n+O(l_p^2)\right),
\end{equation}
whereas \eqref{eq:RD_GDSR} gives
\begin{equation}
E^{\rm (GDSR)}_{n,\pm}=\pm\Omega_n\left(1-\chi l_p\Omega_n+O(l_p^2)\right),
\end{equation}
so the two schemes coincide at first order up to the replacement $l_p\mapsto \chi l_p$. In the nonrelativistic regime ($k_n\ll m$), the positive branch may be written schematically as
\begin{equation}
E_{n,+}=m+\frac{k_n^2}{2m}+\delta E_{\rm def}(m,k_n;l_p,\chi),
\end{equation}
where $\delta E_{\rm def}$ produces a deformation-dependent shift of the rest energy and a deformation-dependent correction to the kinetic term. This highlights that, even in a simple box, Planck-scale modifications can in principle be encoded either as effective mass renormalizations or as higher-order momentum corrections, depending on the adopted map.

\paragraph{Interpretation within the FV framework.}
Because the confinement fixes $k_n$ entirely through boundary conditions on $\Psi$ in configuration space, deformation effects enter only through the stationary relation between $E$ and $k_n$. This separation is particularly transparent in the FV formulation: the two-component structure organizes particle and antiparticle sectors while the boundary-value problem remains formally identical to the nondeformed case at the level of spatial quantization. As a result, the qualitative signatures of the deformation are cleanly identifiable in the spectral branches: MS-type DSR yields bounded energies and eigenvalue accumulation near $\pm 1/l_p$, whereas G-DSR yields unbounded but sublinearly growing energies with decreasing level spacing. These features provide concrete diagnostics for distinguishing deformation scenarios in confined relativistic systems and furnish a controlled setting for subsequent analyses of derived quantities such as level densities, cutoff-sensitive sums, and semiclassical limits.
\section{1D FV oscillator in G-DSR}
\label{sec:KGoscGDSR}

Relativistic oscillator couplings constitute a particularly transparent setting in which to compare the particle and antiparticle branches of relativistic wave equations and to quantify deformation-induced spectral shifts. In the presence of a deformed dispersion relation, the Klein--Gordon (KG) oscillator provides an analytically controllable model where the interplay between non-minimal confinement and Planck-scale corrections can be traced explicitly at the operator level and then propagated to the spectrum in a systematic perturbative expansion \cite{KGO,Moshinsky1989}.

\subsection{FV form of the Klein--Gordon oscillator}
In one spatial dimension, the KG oscillator is introduced through the standard non-minimal substitution
\begin{equation}
\hat p \;\longrightarrow\; \hat p - i m\omega x\,\sigma_3,
\qquad
\hat p=-i\frac{d}{dx},
\eqtag{19}
\end{equation}
which generates harmonic confinement while preserving the relativistic structure of the KG dynamics. Within the
Feshbach--Villars (FV) formalism, the second-order KG equation is recast as a first-order evolution equation for a
two-component wavefunction $\Psi=(\varphi,\chi)^{\mathsf T}$, in which the positive- and negative-frequency sectors are
encoded as dynamical components. The FV Hamiltonian for the free KG field is
\[
\hat H_{FV}=(\sigma_3+i\sigma_2)\,\frac{\hat p^{\,2}}{2m}+m\sigma_3,
\]
and the oscillator coupling yields
\begin{equation}
\hat H^{(\omega)}_{FV}
=(\sigma_3+i\sigma_2)\,
\frac{(\hat p+i m\omega x\,\sigma_3)(\hat p-i m\omega x\,\sigma_3)}{2m}
+m\sigma_3.
\eqtag{20}
\end{equation}

A central feature of the FV representation is the appearance of an \emph{indefinite} metric: the conserved inner
product is
\[
\langle\Psi_1,\Psi_2\rangle_{FV}=\int dx\,\Psi_1^\dagger\,\sigma_3\,\Psi_2,
\]
which distinguishes particle-like and antiparticle-like sectors by the sign of the norm. Consequently, operator
ordering and Hermiticity must be understood with respect to this $\sigma_3$-metric (pseudo-Hermiticity), and any
deformation or coupling should be implemented in a way that preserves consistency of the FV norm and the associated
continuity equation at the order of approximation considered.

\paragraph{a. Operator ordering and FV consistency.}
Because $[\hat p,x]\neq 0$, the oscillator substitution must be implemented as the ordered product in Eq.~(20), rather
than as a naive square. This is not merely a technicality: in FV form, the kinetic operator is multiplied by
$(\sigma_3+i\sigma_2)$, and the ordering-induced term propagates into the diagonal/off-diagonal structure that couples
$\varphi$ and $\chi$, thereby affecting both the branch structure and the normalization in the $\sigma_3$ inner product.

To make the ordering contribution explicit, we use
\[
(\hat p+i m\omega A)(\hat p-i m\omega A)=\hat p^{\,2}+m^2\omega^2A^2+i m\omega[\hat p,A].
\]
Choosing $A=x$ and using $[\hat p,x]=-i$, we obtain
\begin{equation}
(\hat p+i m\omega x)(\hat p-i m\omega x)=\hat p^{\,2}+m^2\omega^2x^2+m\omega.
\eqtag{20a}
\end{equation}
Since $\sigma_3$ commutes with both $\hat p$ and $x$, it does not modify the commutator itself; it only appears as a
matrix factor in the constant term. Hence,
\begin{equation}
(\hat p+i m\omega x\,\sigma_3)(\hat p-i m\omega x\,\sigma_3)
=\hat p^{\,2}+m^2\omega^2x^2+m\omega\,\sigma_3.
\eqtag{20b}
\end{equation}
Equation (20b) shows that the correct ordering generates an additional constant contribution (in the FV matrix sense),
which would be missed by replacing the ordered product with a naive square. This term is ultimately responsible for the
$(2n+1)$ shift in the relativistic spectrum and is therefore essential when comparing undeformed and deformed branches
on equal footing.

\paragraph{b. Oscillator eigenproblem.}
The scalar harmonic-oscillator operator extracted from Eq.~(20b) is
\begin{equation}
\left[
-\frac{d^2}{dx^2}+m^2\omega^2x^2
\right]\phi_n(x)=\Lambda_n\,\phi_n(x),
\qquad
\Lambda_n=m\omega\,2n,
\qquad
n=0,1,2,\dots
\eqtag{21}
\end{equation}
where $\phi_n$ are the usual Hermite functions. In the FV setting these functions provide the natural spatial basis in
which the two-component FV wavefunction can be expanded, while the branch information resides in the internal
($\sigma$-matrix) structure.

\paragraph{c. Undeformed relativistic spectrum and branch pairing.}
With the ordering in Eq.~(20) (equivalently Eq.~(20b)), the commutator term shifts the effective eigenvalue entering the
decoupled FV equation. It is therefore convenient to define
\begin{equation}
\Xi_n \equiv \Lambda_n + m\omega = m\omega(2n+1),
\qquad n=0,1,2,\dots
\eqtag{22a}
\end{equation}
so that the undeformed energy branches ($l_p\to 0$) take the form
\begin{equation}
\big(E^{(0)}_{n,\pm}\big)^2 = m^2+\Xi_n = m^2+m\omega(2n+1),
\qquad
E^{(0)}_{n,\pm}=\pm\sqrt{m^2+m\omega(2n+1)}.
\eqtag{22}
\end{equation}
In FV language, the $\pm$ signs correspond to states of opposite $\sigma_3$-norm (particle/antiparticle sectors), and
the symmetry $E\mapsto -E$ is reflected in the pseudo-Hermitian structure of $\hat H_{FV}$ with respect to the
$\sigma_3$ metric.

\subsection{First-order G-DSR correction as an energy-dependent FV perturbation}
To leading order in the Planck length $l_p$, the G-DSR Hamiltonian may be organized as a perturbative deformation of the
FV oscillator Hamiltonian,
\begin{equation}
\hat H^{(\omega)}_{G\text{-}DSR}=\hat H^{(\omega)}_{FV}+l_p\,\delta\hat H^{(\omega)}+O(l_p^2),
\eqtag{23pre}
\end{equation}
where, for stationary solutions, the energy $E$ is treated as a c-number parameter. In the FV setting this point is
crucial: the deformation becomes an \emph{energy-dependent} operator, so first-order corrections must be computed using
FV perturbation theory with the $\sigma_3$-weighted inner product. This guarantees that the correction is compatible
with the FV norm and with the particle/antiparticle interpretation encoded by $\sigma_3$.

Using the generic deformation structure in FV form and implementing the oscillator coupling with the same ordering as
in Eq.~(20), the leading correction can be written as
\begin{equation}
\delta \hat H^{(\omega)}
=
\alpha_2 E^2\,\sigma_3
-\Delta\alpha\,\frac{E}{m}\,
(\hat p+i m\omega x\,\sigma_3)(\hat p-i m\omega x\,\sigma_3)\,
(\sigma_3+i\sigma_2),
\eqtag{23}
\end{equation}
with $\Delta\alpha\equiv \alpha_3-\alpha_1$. The ordered product is required for internal consistency: it ensures that
the same commutator-induced term that fixes the undeformed spectrum also appears inside the deformation operator, so
that the $O(l_p)$ shift is not contaminated by an inconsistent treatment of ordering.

\paragraph{a. FV first-order energy shift.}
The first-order correction is obtained from FV perturbation theory,
\begin{equation}
\Delta E_{n,\pm}
=
l_p\,
\frac{\langle n,\pm|\sigma_3\,\delta \hat H^{(\omega)}|n,\pm\rangle}
{\langle n,\pm|\sigma_3|n,\pm\rangle}
+O(l_p^2),
\eqtag{24}
\end{equation}
where the $\sigma_3$ insertions implement the FV metric. At this order, the required matrix elements reduce to harmonic
oscillator expectation values of the quadratic operator in Eq.~(20b). For compactness, we introduce
\begin{equation}
\lambda_n \equiv \Xi_n = m\omega(2n+1),
\qquad n=0,1,2,\dots
\eqtag{25}
\end{equation}
which explicitly includes the ordering-induced shift. The perturbed spectrum can then be written as
\begin{equation}
E_{n,\pm}
\simeq
E^{(0)}_{n,\pm}
+
l_p\left[
\alpha_2\big(E^{(0)}_{n,\pm}\big)^2-\Delta\alpha\,\lambda_n
\right]\frac{E^{(0)}_{n,\pm}}{m}
+O(l_p^2).
\eqtag{26}
\end{equation}
The perturbative regime requires $l_p|E^{(0)}_{n,\pm}|\ll 1$, which for oscillator levels becomes
$l_p\sqrt{m^2+\lambda_n}\ll 1$. In FV terms, this ensures that the deformation does not invalidate the separation into
weakly corrected particle/antiparticle branches and that the $\sigma_3$-norm assignment remains stable at $O(l_p)$.

\subsection{Extended AC and MS analyses}

\subsubsection*{1. AC-type deformation $(\alpha_2=0,\ \Delta\alpha=1)$}
For the Amelino--Camelia (AC) choice, Eq.~(26) reduces to a purely level-dependent correction,
\begin{equation}
E^{(AC)}_{n,\pm}
\simeq
E^{(0)}_{n,\pm}
-
l_p\,\lambda_n\,
\frac{E^{(0)}_{n,\pm}}{m}
+O(l_p^2).
\eqtag{27}
\end{equation}
At fixed $(m,\omega)$ the magnitude of the shift increases with $n$ through $\lambda_n$, so the deformation becomes more
pronounced for highly excited states (subject to $l_p|E|\ll 1$). From the FV perspective, the correction scales with the
expectation value of the ordered kinetic-plus-confining operator, i.e.\ with the same combination that controls the
relative weight of the two FV components $(\varphi,\chi)$ as one moves to higher levels.

Within the perturbative window, one may interpret the deformation semiclassically as an effective, mildly
energy-dependent reduction of the oscillator scale,
\begin{equation}
\omega \;\longrightarrow\; \omega^{(AC)}_{\mathrm{eff}}(E)
\simeq \omega\,(1-l_pE),
\eqtag{28}
\end{equation}
so that the level spacing decreases as $|E|$ increases.

\subsubsection*{2. MS-type deformation $(\alpha_2=1,\ \Delta\alpha=1)$}
For the Magueijo--Smolin (MS) choice, the mass-shell contribution competes with the level-dependent term, yielding
\begin{equation}
E^{(MS)}_{n,\pm}
\simeq
E^{(0)}_{n,\pm}
+
l_p\left[
\big(E^{(0)}_{n,\pm}\big)^2-\lambda_n
\right]\frac{E^{(0)}_{n,\pm}}{m}
+O(l_p^2).
\eqtag{29}
\end{equation}
Relative to AC-type models, MS-type corrections are typically more sensitive in the ultra-relativistic regime because
the term proportional to $\big(E^{(0)}\big)^2$ becomes dominant at large energies (still within $l_p|E|\ll 1$). In FV
language, this reflects the fact that the deformation acts partly through the diagonal $\sigma_3$ sector, directly
modifying the relative placement of the two frequency branches rather than only renormalizing the level-dependent
operator.

Equivalently, the MS correction may be cast as an energy-dependent renormalization of the inertial mass,
\begin{equation}
m \;\longrightarrow\; m^{(MS)}_{\mathrm{eff}}(E)\simeq m\,(1+l_pE),
\eqtag{30}
\end{equation}
with corresponding modifications of the oscillator scales.

\subsubsection*{3. Branch structure and particle--antiparticle pairing}
Because the FV inner product distinguishes the two energy branches through the $\sigma_3$ metric, the first-order
corrections preserve particle--antiparticle pairing whenever the undeformed spectrum is symmetric under $E\to -E$. In
particular, Eq.~(26) implies
\begin{equation}
\Delta E_{n,-}=-\Delta E_{n,+}
\qquad \text{at fixed }|E^{(0)}_{n,\pm}|,
\eqtag{31}
\end{equation}
so charge-conjugation pairing is maintained at $O(l_p)$, and the FV interpretation of densities and currents remains
consistent at this order.
\section{Tunneling in the Feshbach--Villars formalism: SR, DSR, and G-DSR frameworks}
\label{sec:tunneling_FV}

Barrier scattering and tunneling constitute sensitive probes of flux conservation and of how modified dispersion
relations (MDRs) reshape wave numbers, group velocities, and conserved currents in relativistic quantum dynamics
\cite{GreinerRQM,HaugeStovneng1989}. In the Feshbach--Villars (FV) representation of the Klein--Gordon equation, these
effects can be tracked explicitly because the dynamics is cast in a Schr\"odinger-like form with a conserved (but
indefinite) inner product and an associated conserved current. Throughout this section we use natural units
$\hbar=c=1$.

\subsection{FV setup and rectangular barrier}
We consider a one-dimensional electrostatic barrier of height $V_0$ and width $a$,
\begin{equation}
V(x)=
\begin{cases}
0, & x<0,\\
V_0, & 0\le x \le a,\\
0, & x>a,
\end{cases}
\label{eq:barrierV_FV}
\end{equation}
and a scalar particle of mass $m$ incident from the left. In the FV representation the two-component wave function
$\Psi(x,t)$ satisfies
\begin{equation}
i\frac{\partial}{\partial t}\Psi(x,t)=\hat H_{FV}[V]\Psi(x,t),
\label{eq:FV_Schro_barrier}
\end{equation}
where $\hat H_{FV}[V]$ is the FV Hamiltonian in the presence of the electrostatic potential. For stationary states,
\begin{equation}
\Psi(x,t)=e^{-iEt}\Psi(x),
\label{eq:stationary_ansatz_FV}
\end{equation}
the FV equation reduces to the eigenvalue problem $\hat H_{FV}[V]\Psi=E\Psi$.

In each region where $V(x)$ is constant, the FV system decouples into the standard stationary Klein--Gordon form for a
scalar mode $\psi$,
\begin{equation}
\frac{d^2\psi}{dx^2}+\Big[(E-V)^2-m^2\Big]\psi=0,
\label{eq:KG_region_FV}
\end{equation}
so the spatial dependence is controlled by the local wave number
\begin{equation}
k(V)=\sqrt{(E-V)^2-m^2}.
\label{eq:k_local}
\end{equation}
In regions I and III ($V=0$) we write
\begin{equation}
k\equiv \sqrt{E^2-m^2},
\label{eq:k_out}
\end{equation}
and inside the barrier (region II, $V=V_0$),
\begin{equation}
q\equiv \sqrt{(E-V_0)^2-m^2}.
\label{eq:q_in}
\end{equation}
In the evanescent (tunneling) regime,
\begin{equation}
(E-V_0)^2<m^2
\quad\Longrightarrow\quad
q=i\kappa,
\qquad
\kappa\equiv\sqrt{m^2-(E-V_0)^2}.
\label{eq:kappa_FV}
\end{equation}

\subsection{FV current and definition of reflection and transmission}
In the FV formalism the conserved density and current are
\begin{equation}
\rho_{FV}=\Psi^\dagger\sigma_3\Psi,
\qquad
j_{FV}=\frac{1}{2mi}\left[\Psi^\dagger(\sigma_3+i\sigma_2)\,\partial_x\Psi
-(\partial_x\Psi^\dagger)(\sigma_3+i\sigma_2)\Psi\right],
\label{eq:FV_current}
\end{equation}
so that $\partial_t\rho_{FV}+\partial_x j_{FV}=0$. In the asymptotic free regions, plane-wave solutions carry a constant
FV flux. We therefore define the reflection and transmission coefficients (we use the standard notation: $R$ for
reflection and $T$ for transmission)
\begin{equation}
R\equiv \frac{|j_{\rm ref}|}{j_{\rm inc}},
\qquad
T\equiv \frac{j_{\rm tr}}{j_{\rm inc}},
\label{eq:RT_def_FV}
\end{equation}
so that, in the subcritical regime where no pair-production channels are activated, FV current conservation implies
$R+T=1$.

\subsection{Standard (SR) FV barrier transmission}
In regions I ($x<0$), II ($0<x<a$), and III ($x>a$), we write the scalar modes as
\begin{align}
\psi_I(x) &= e^{ikx}+r\,e^{-ikx},\\
\psi_{II}(x) &= A\,e^{\kappa x}+B\,e^{-\kappa x},\\
\psi_{III}(x) &= t\,e^{ikx},
\end{align}
where $k$ and $\kappa$ are given by \eqref{eq:k_out} and \eqref{eq:kappa_FV}. Matching $\psi$ and $d\psi/dx$ at $x=0$
and $x=a$ yields the standard transfer-matrix result. In the tunneling regime $q=i\kappa$,
\begin{equation}
T_{\rm std}(E)=
\frac{1}{1+\displaystyle
\frac{(k^2+\kappa^2)^2}{4k^2\kappa^2}\sinh^2(\kappa a)} ,
\label{eq:Tstd_FV}
\end{equation}
and
\begin{equation}
R_{\rm std}(E)=1-T_{\rm std}(E).
\label{eq:Rstd_FV}
\end{equation}
These expressions apply for propagating incidence ($E>m$ so that the incident FV flux is nonvanishing) and in a
subcritical domain where Klein-paradox effects are neglected.

\subsection{Standard DSR in the FV framework}
To incorporate a standard DSR deformation in the FV scattering problem, we keep the matching conditions unchanged and
implement the MDR through an effective-energy map. A typical MS-like choice is
\begin{equation}
E \longrightarrow E_{\rm DSR}=\frac{E}{1-l_p E},
\qquad
E-V_0 \longrightarrow (E-V_0)_{\rm DSR}=\frac{E-V_0}{1-l_p(E-V_0)}.
\label{eq:EdSR_FV}
\end{equation}
The corresponding wave numbers in each region become
\begin{equation}
k_{\rm DSR}(l_p)=\sqrt{E_{\rm DSR}^2-m^2},
\qquad
q_{\rm DSR}(l_p)=\sqrt{(E-V_0)_{\rm DSR}^2-m^2},
\label{eq:kqDSR_FV}
\end{equation}
with $q_{\rm DSR}=i\kappa_{\rm DSR}$ in the tunneling regime and
$\kappa_{\rm DSR}(l_p)=\sqrt{m^2-(E-V_0)_{\rm DSR}^2}$.
Replacing $(k,\kappa)$ by $(k_{\rm DSR},\kappa_{\rm DSR})$ in \eqref{eq:Tstd_FV} yields
\begin{equation}
T_{\rm DSR}(E;l_p)=
\frac{1}{1+\displaystyle
\frac{\big(k_{\rm DSR}^2+\kappa_{\rm DSR}^2\big)^2}{4k_{\rm DSR}^2\kappa_{\rm DSR}^2}
\sinh^2\!\big(\kappa_{\rm DSR}\,a\big)} ,
\label{eq:Tdsr_FV}
\end{equation}
and $R_{\rm DSR}(E;l_p)=1-T_{\rm DSR}(E;l_p)$.
The mapping \eqref{eq:EdSR_FV} is well defined only when $1-l_pE>0$ and $1-l_p(E-V_0)>0$, thereby avoiding the
singularities characteristic of rational DSR parametrizations.

\subsection{Generalized DSR (G-DSR) in the FV framework}
In the generalized DSR (G-DSR) setting, we implement a flexible first-order deformation consistent with FV perturbative
structures via
\begin{equation}
E \longrightarrow E_{\rm eff}=E\left(1+\chi\,l_p E\right),
\qquad
E-V_0 \longrightarrow (E-V_0)_{\rm eff}=(E-V_0)\left(1+\chi\,l_p(E-V_0)\right),
\label{eq:Eeff_FV}
\end{equation}
where $\chi$ is a dimensionless model coefficient. The associated wave numbers are
\begin{equation}
k_{\rm GDSR}(l_p)=\sqrt{E_{\rm eff}^2-m^2},
\qquad
q_{\rm GDSR}(l_p)=\sqrt{(E-V_0)_{\rm eff}^2-m^2},
\label{eq:kqGDSR_FV}
\end{equation}
with $q_{\rm GDSR}=i\kappa_{\rm GDSR}$ in the tunneling regime and
$\kappa_{\rm GDSR}(l_p)=\sqrt{m^2-(E-V_0)_{\rm eff}^2}$.
Replacing $(k,\kappa)$ by $(k_{\rm GDSR},\kappa_{\rm GDSR})$ in \eqref{eq:Tstd_FV} yields $T_{\rm GDSR}(E;l_p)$, and
$R_{\rm GDSR}(E;l_p)=1-T_{\rm GDSR}(E;l_p)$.

\subsection{Numerical illustration: joint plots of $R(E)$ and $T(E)$}
For a representative configuration we take
\begin{equation}
m=1,\qquad V_0=2,\qquad a=4,\qquad E\in[0,10],
\qquad l_p\in\{0,\ 0.02,\ 0.06\},
\qquad \chi=1,
\label{eq:numparams_Escan_FV}
\end{equation}
and evaluate $R(E)$ and $T(E)$ for the SR, DSR, and G-DSR prescriptions. To facilitate direct comparison of flux
partition, we plot $R$ and $T$ on the same axes for each framework; solid curves denote $T(E)$ and dashed curves denote
$R(E)$.

\begin{figure}[t]
\centering
\safeincludegraphics[width=0.82\linewidth]{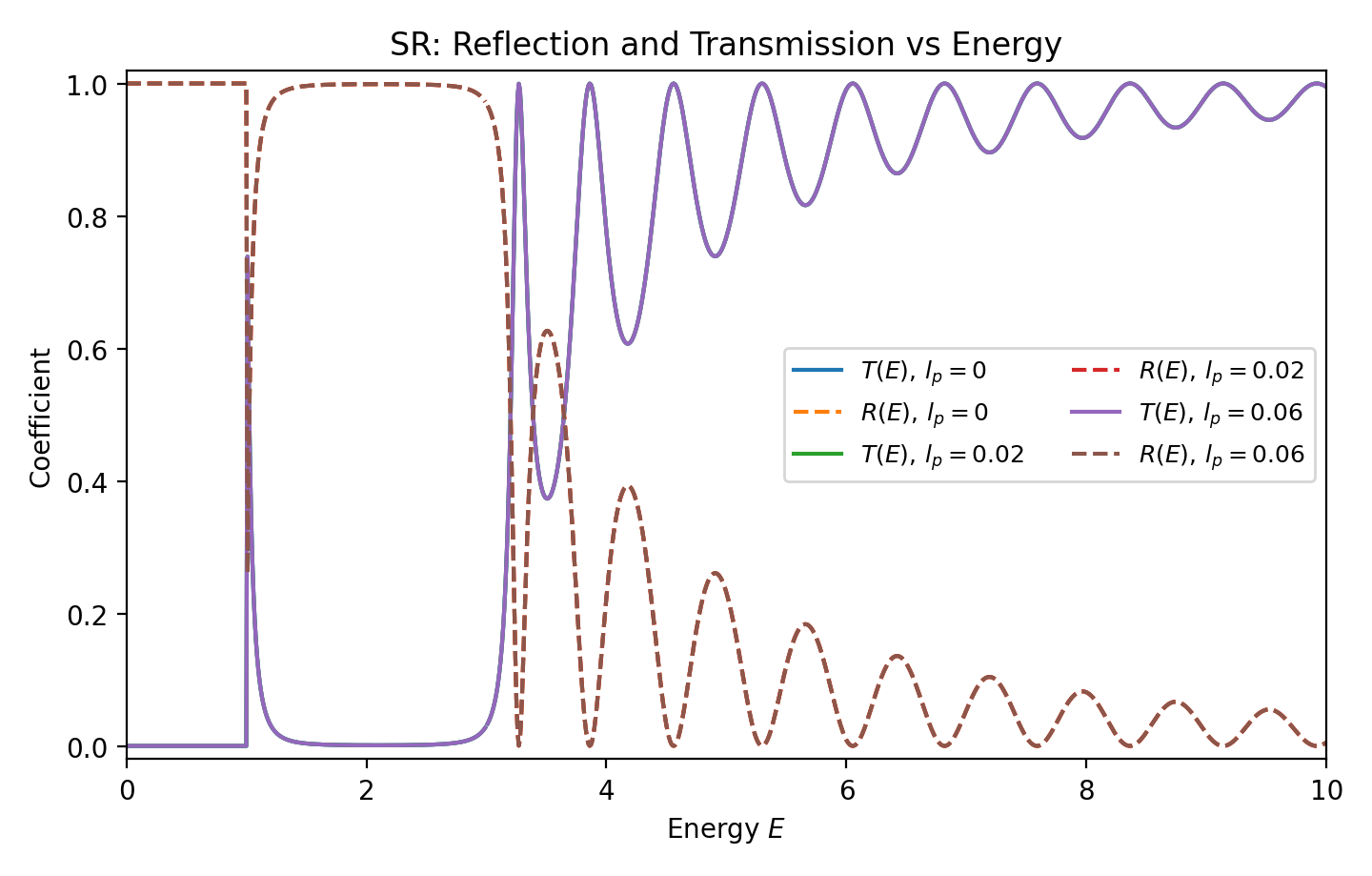}
\caption{Standard (SR) case: transmission $T(E)$ (solid) and reflection $R(E)$ (dashed) in the FV formalism for
$l_p\in\{0,0.02,0.06\}$. (For SR, the curves coincide because there is no deformation.) Parameters are given in
\eqref{eq:numparams_Escan_FV}.}
\label{fig:SR_RT_FV}
\end{figure}

\begin{figure}[t]
\centering
\safeincludegraphics[width=0.82\linewidth]{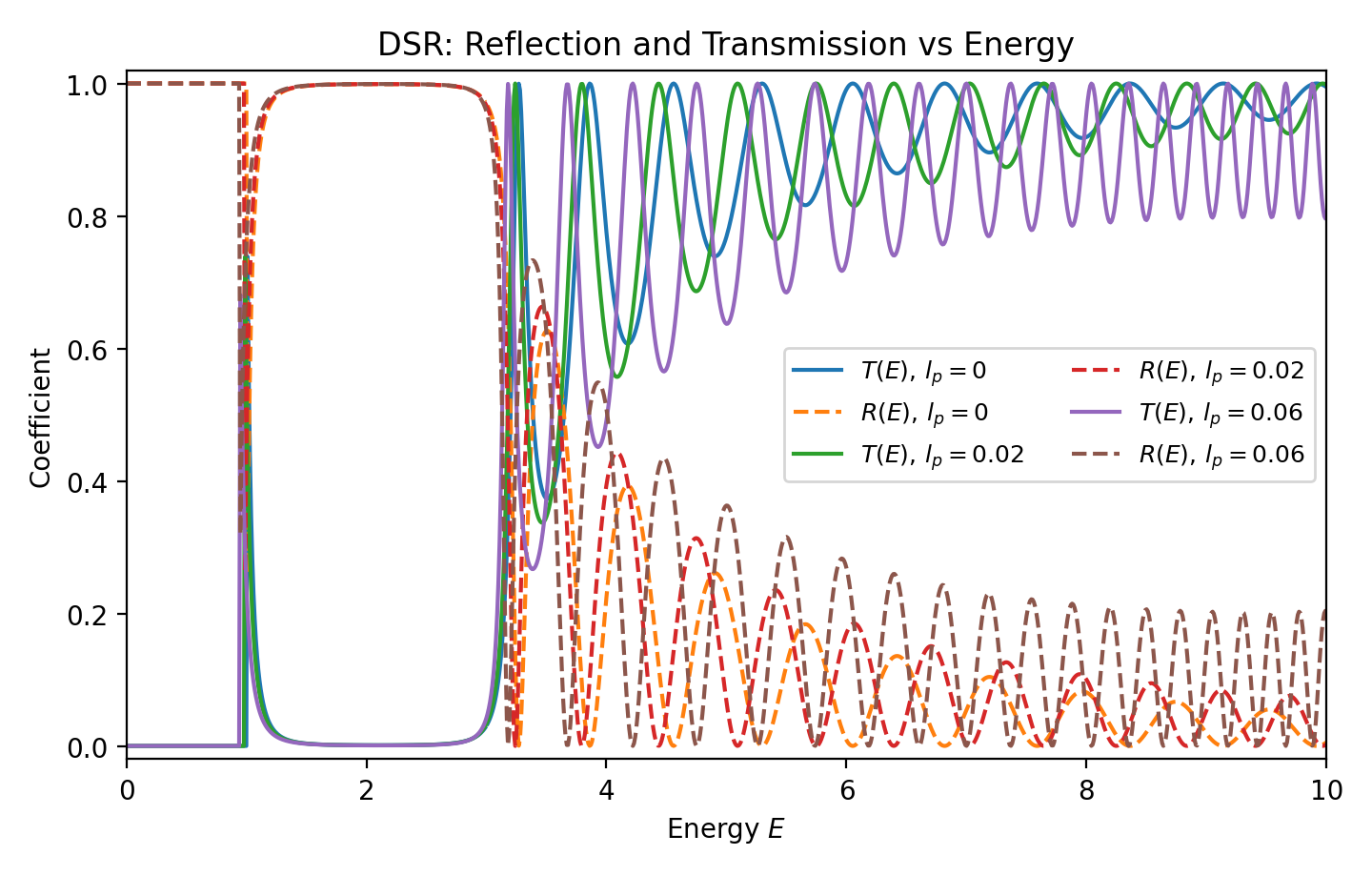}
\caption{Standard DSR case: transmission $T(E)$ (solid) and reflection $R(E)$ (dashed) for $l_p\in\{0,0.02,0.06\}$,
using the FV flux definition \eqref{eq:RT_def_FV} and parameters \eqref{eq:numparams_Escan_FV}.}
\label{fig:DSR_RT_FV}
\end{figure}

\begin{figure}[t]
\centering
\safeincludegraphics[width=0.82\linewidth]{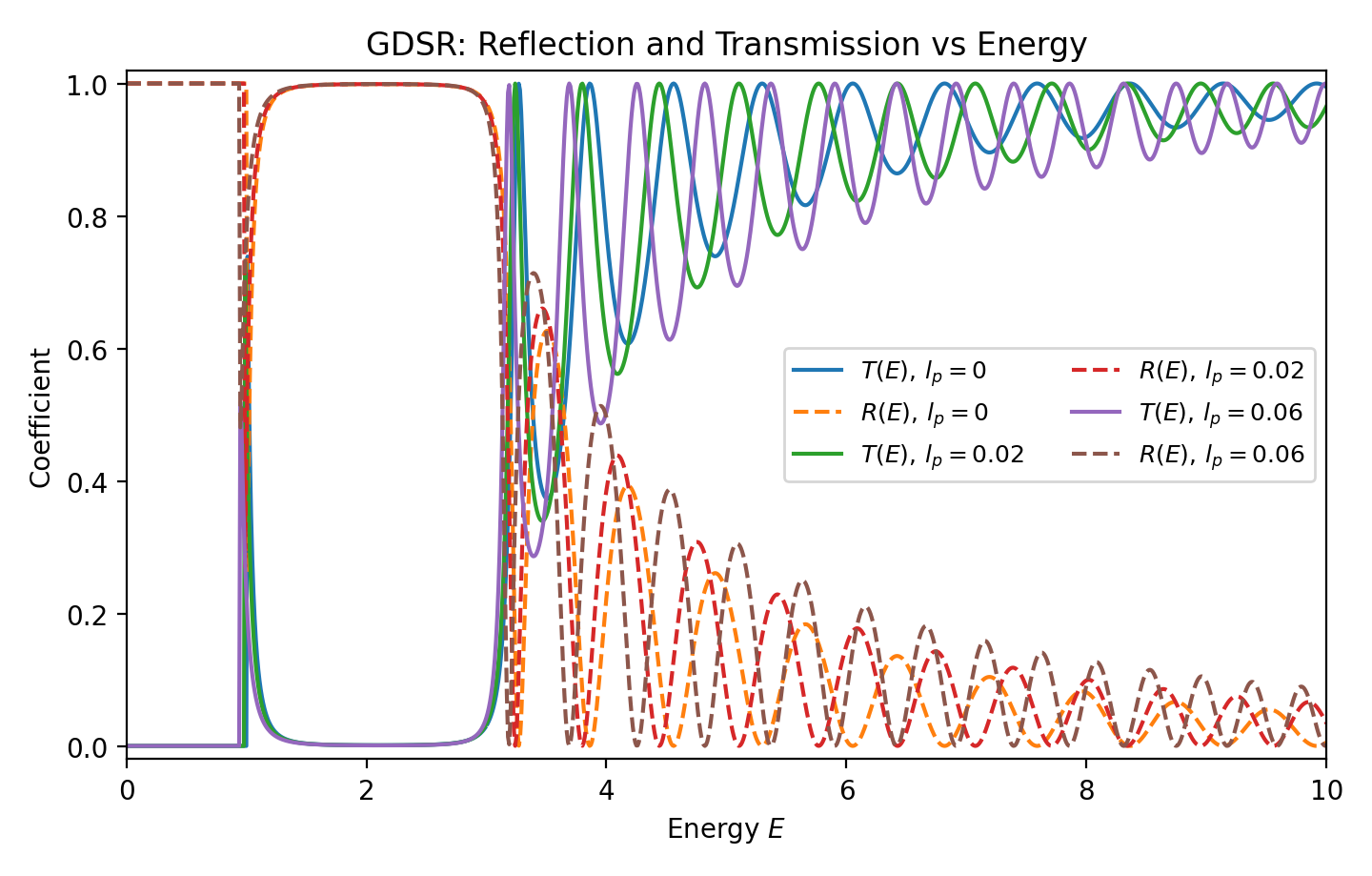}
\caption{G-DSR case: transmission $T(E)$ (solid) and reflection $R(E)$ (dashed) for $l_p\in\{0,0.02,0.06\}$, with
$\chi=1$ and parameters \eqref{eq:numparams_Escan_FV}.}
\label{fig:GDSR_RT_FV}
\end{figure}

\paragraph{Extended discussion.}
\emph{(i) FV flux conservation and the role of the indefinite metric.}
The FV current \eqref{eq:FV_current} is constructed to satisfy a continuity equation with respect to the $\sigma_3$
metric. In subcritical scattering, the asymptotic FV flux is conserved, and the definitions
\eqref{eq:RT_def_FV} enforce $R(E)+T(E)=1$ for propagating incidence ($E>m$). This provides a stringent internal
consistency check on any MDR implementation: modifying the dispersion while leaving boundary conditions unchanged must
still yield a conserved asymptotic FV current at the working order.

\emph{(ii) Why MDR effects primarily enter through the tunneling exponent.}
In the evanescent domain, the transmission is exponentially controlled by $\kappa a$ (or its deformed counterpart),
since $T$ depends on $\sinh^2(\kappa a)$ in \eqref{eq:Tstd_FV}. Consequently, small $l_p$-induced changes in the
effective energy inside the barrier can produce comparatively large changes in $T(E)$ when $\kappa a\gtrsim 1$. This
explains why increasing the barrier width $a$ (or moving deeper into the tunneling regime) amplifies Planck-scale
sensitivities even when $l_p E\ll 1$.

\emph{(iii) DSR versus G-DSR: rational versus polynomial deformations.}
The standard DSR map \eqref{eq:EdSR_FV} is rational and introduces a model-imposed domain of validity through the
conditions $1-l_pE>0$ and $1-l_p(E-V_0)>0$. Within this domain, it can significantly reshape the effective wave numbers
by enhancing $E_{\rm DSR}$ relative to $E$ as $E$ increases, thereby modifying both the above-barrier oscillatory
behavior (through $q_{\rm DSR}$ real) and the below-barrier suppression (through $\kappa_{\rm DSR}$). By contrast, the
G-DSR prescription \eqref{eq:Eeff_FV} is polynomial and is naturally interpreted as a controlled first-order expansion:
it avoids poles and organizes MDR corrections in a way that is directly compatible with FV perturbative treatments.

\emph{(iv) Energy thresholds and interpretation of the low-energy region.}
The FV definition of $R$ and $T$ presupposes a nonzero incident flux, which requires a propagating asymptotic mode in
region I, i.e.\ $E>m$ (or, in deformed models, the corresponding condition on the effective energy). For $E\le m$, the
asymptotic wave number becomes imaginary and the usual scattering interpretation is not applicable; in practice one
restricts the physical discussion of tunneling coefficients to the propagating domain.

\emph{(v) Above-barrier regime and resonance structure.}
When the interior wave number is real (above-barrier propagation), the barrier supports interference resonances that
manifest as oscillations in $T(E)$ and complementary oscillations in $R(E)$. MDRs shift these structures by changing
the phase $q(l_p)a$. Hence, the deformation does not merely renormalize the overall transmission level; it also
displaces resonance conditions in energy, providing an additional diagnostic beyond exponential tunneling suppression.

\section{Discussion}
A central advantage of the Feshbach--Villars representation is that it reorganizes Klein--Gordon dynamics into a two-component, Schr\"odinger-like evolution while retaining a conserved quantity defined with respect to an indefinite metric. In this framework, the fundamental structural requirement for conservation laws is \(\sigma_3\)-pseudo-Hermiticity of the Hamiltonian. In the present construction, this property is preserved at \(O(l_p)\) by adopting a consistent symmetrized ordering for mixed energy--momentum contributions induced by the MDR, thereby maintaining the FV continuity equation and the standard expression for the FV current at the working order. Consequently, the scattering flux accounting remains internally consistent, and in stationary problems the current-based definitions of reflection and transmission continue to enforce flux partitioning (e.g., \(R+T=1\) in the subcritical regime), providing a stringent diagnostic for the correctness of the deformation prescription.

From a physical standpoint, the generalized leading-order MDR introduces two qualitatively distinct mechanisms. The coefficient \(\alpha_2\) generates an energy-dependent modification of the mass shell, whereas \(\Delta\alpha\) deforms the effective kinetic sector and therefore alters the relation between local wave number, group velocity, and FV current. In bound-state settings such as the Klein--Gordon oscillator, these mechanisms can compete: kinetic deformations naturally act to compress the spectrum by reducing effective level spacing, while mass-shell deformations shift both FV branches in an explicitly energy-dependent manner. In scattering, the same interplay is encoded in the deformed wave numbers and manifests in both the magnitude and sign structure of transmitted FV flux, which is particularly relevant in the onset and development of the supercritical (pair-production) regime.

The analysis also highlights limitations intrinsic to a leading-order treatment. First, the results are quantitatively reliable only within the perturbative domain \(l_p E \ll 1\) (and, for step or barrier backgrounds, \(l_p |E-V_0| \ll 1\)), where truncation at \(O(l_p)\) is justified. Second, the momentum-space map employed here should be regarded as an effective parametrization; more microscopic realizations may generate higher-order corrections and additional operator-ordering subtleties, especially in spatially varying external fields. Third, in the deep supercritical regime a physically complete description requires a quantum-field-theoretic treatment of pair production and vacuum polarization; the one-particle FV approach captures the kinematic signature of supercriticality and ensures flux conservation at the level of the deformed current, but it does not incorporate backreaction or vacuum dynamics.

Within these caveats, a robust qualitative conclusion emerges: Planck-scale deformations can operate as an effective regulator of supercritical scattering. In particular, MS-type realizations tend to shift the supercritical threshold upward and diminish the magnitude of negative transmitted flux, trends that favor vacuum stability in the perturbative regime. More broadly, the results indicate that the momentum-space geometry encoded by the map parameters controls both when the supercritical regime sets in and how strongly it is expressed in observable reflection/transmission characteristics.

\paragraph{FV perspectives.}
From the viewpoint of the FV formalism, several extensions are particularly natural. First, \(\sigma_3\)-pseudo-Hermiticity offers a practical selection criterion for admissible operator orderings and couplings when MDRs are implemented in external fields or in nontrivial backgrounds, where ordering ambiguities can otherwise spoil conservation laws. Second, FV methods are well suited to curved and topologically nontrivial spacetimes, where a conserved current is essential to define scattering and spectral observables; combining this with DSR-inspired deformations provides a controlled avenue to study Planck-scale kinematics beyond flat space. Finally, FV provides a useful bridge between first-quantized calculations and field-theoretic interpretations: the onset of regimes where a single-particle picture becomes insufficient is sharply flagged by current signatures, helping to delineate where a second-quantized treatment of pair production is unavoidable.
\section{Conclusion}
We have established a consistent first-order Feshbach--Villars formulation for Klein--Gordon dynamics governed by a generic leading-order, DSR-modified dispersion relation. By constructing the deformed FV Hamiltonian so that \(\sigma_3\)-pseudo-Hermiticity is preserved at \(O(l_p)\), we guarantee conservation of the FV charge and current and thereby provide a clear proof of flux conservation for stationary scattering within the validity of the perturbative expansion. This pseudo-Hermitian foundation is especially valuable because it permits a branch-resolved interpretation of particle/antiparticle sectors while keeping scattering observables anchored to a conserved current defined by the FV metric.

As applications, we analyzed two paradigmatic problems. For the one-dimensional Klein--Gordon oscillator, we derived explicit \(O(l_p)\) corrections to the positive- and negative-energy branches and clarified how kinetic versus mass-shell deformations reshape the level spacing and the high-energy spectral density. For the electrostatic step potential, we obtained deformation-dependent wave numbers, computed reflection and transmission coefficients directly from the FV current, and quantified the displacement of the supercritical threshold associated with the supercritical regime. A systematic comparison of Amelino--Camelia and Magueijo--Smolin realizations indicates that MS-type deformations more efficiently postpone the onset of the supercritical regime and suppress the magnitude of the negative transmitted flux, suggesting a perturbative tendency toward regulating supercritical scattering.

In addition to higher-order corrections in \(l_p\), an important direction is to extend the present construction to curved or defect spacetimes, where the FV current remains a natural anchor for defining physically meaningful scattering and spectral observables. In that setting, pseudo-Hermiticity can be used as a guiding principle to build consistent deformed Hamiltonians that retain conservation laws while incorporating both background geometry and Planck-scale kinematics.

Several natural extensions follow. These include higher-order calculations in \(l_p\), alternative momentum-space maps and composition laws, and a fully field-theoretic treatment of pair production in deformed kinematics. In all such developments, FV pseudo-Hermiticity provides a principled criterion for preserving consistent conservation laws and for maintaining a physically meaningful definition of flux-based observables.

\bibliographystyle{apalike}
\bibliography{references}

\begin{thebibliography}{}

\bibitem[Amelino-Camelia, 2001]{AC2001}
Amelino-Camelia, G. (2001).
\newblock Testable scenario for relativity with minimum length.
\newblock {\em Physics Letters B}, 510(1--4):255--263.

\bibitem[Amelino-Camelia, 2002a]{ACopen}
Amelino-Camelia, G. (2002a).
\newblock Doubly-special relativity: First results and key open problems.
\newblock {\em International Journal of Modern Physics D}, 11(10):1643--1669.

\bibitem[Amelino-Camelia, 2002b]{AC}
Amelino-Camelia, G. (2002b).
\newblock Relativity in spacetimes with short-distance structure governed by an
  observer-independent (planckian) length scale.
\newblock {\em International Journal of Modern Physics D}, 11(1):35--59.

\bibitem[Amelino-Camelia et~al., 2003]{ACphen2003}
Amelino-Camelia, G., Kowalski-Glikman, J., Mandanici, G., and Procaccini, A.
  (2003).
\newblock Phenomenology of doubly special relativity.
\newblock {\em International Journal of Modern Physics A}, 20(34):6007--6038.

\bibitem[Boumali et~al., 2025]{BoumaliJafariEPJC2025}
Boumali, A., Jafari, N., Shukirgaliyev, B., and Serdouk, F. (2025).
\newblock Thermal properties of klein--gordon oscillator in the context of
  amelino-camelia and magueijo--smolin doubly special relativity (dsr)
  frameworks.
\newblock {\em The European Physical Journal C}.

\bibitem[Bouzenada and Boumali, 2023]{BouzenadaBoumali2023}
Bouzenada, A. and Boumali, A. (2023).
\newblock Statistical properties of the two dimensional feshbach--villars
  oscillator (fvo) in the rotating cosmic string space-time.
\newblock {\em Annals of Physics}, 452:169302.

\bibitem[Bouzenada et~al., 2023]{NPBKK2023}
Bouzenada, A., Boumali, A., Vit{\'o}ria, R. L.~L., Ahmed, F., and Al-Raeei, M.
  (2023).
\newblock Feshbach--villars oscillator in {K}aluza--{K}lein theory.
\newblock {\em Nuclear Physics B}, 994:116288.

\bibitem[Bruce and Minning, 1993]{KGO}
Bruce, S. and Minning, P. (1993).
\newblock The klein--gordon oscillator.
\newblock {\em Il Nuovo Cimento A}, 106(5):711--713.

\bibitem[Carmona et~al., 2021]{CarmonaRelancio2021}
Carmona, J.~M., Cort{\'e}s, J.~L., and Relancio, J.~J. (2021).
\newblock Curved momentum space, locality, and generalized space-time.
\newblock {\em Universe}, 7(4):99.

\bibitem[Chargui et~al., 2022]{Chargui2022}
Chargui, Y., Dhahbi, A., and Karam, A.~R. (2022).
\newblock Scattering of relativistic spinless particles within the
  feshbach--villars formalism.
\newblock {\em Heliyon}, 8(10):e11215.

\bibitem[Feshbach and Villars, 1958]{FV}
Feshbach, H. and Villars, F. (1958).
\newblock Elementary relativistic wave mechanics of spin 0 and spin 1/2
  particles.
\newblock {\em Reviews of Modern Physics}, 30(1):24--45.

\bibitem[Fuda, 1980]{Fuda1980}
Fuda, M.~G. (1980).
\newblock Feshbach--villars formalism and pion--nucleon scattering.
\newblock {\em Physical Review C}, 21(4):1480--1491.

\bibitem[Garah and Boumali, 2025]{GarahBoumali2025}
Garah, S. and Boumali, A. (2025).
\newblock Quantum dynamics of scalar particles in a spinning cosmic string
  background with topological defects: A feshbach--villars formalism
  perspective.
\newblock {\em The European Physical Journal C}, 85:1257.

\bibitem[Greiner, 2000]{GreinerRQM}
Greiner, W. (2000).
\newblock {\em Relativistic Quantum Mechanics: Wave Equations}.
\newblock Springer, Berlin, Heidelberg, 3 edition.

\bibitem[Hauge and St{\o}vneng, 1989]{HaugeStovneng1989}
Hauge, E.~H. and St{\o}vneng, J.~A. (1989).
\newblock Tunneling times: A critical review.
\newblock {\em Reviews of Modern Physics}, 61(4):917--936.

\bibitem[Jafari, 2024]{Jafari2024}
Jafari, N. (2024).
\newblock Dsr transformations with zero time delay.

\bibitem[Jafari and Boumali, 2025]{BoumaliJafariPLB2025}
Jafari, N. and Boumali, A. (2025).
\newblock Dirac oscillator in dsr: A comparative study of magueijo--smolin and
  amelino--camelia models.
\newblock {\em Physics Letters B}.

\bibitem[Kowalski-Glikman, 2005]{KowalskiGlikman2005}
Kowalski-Glikman, J. (2005).
\newblock Introduction to doubly special relativity.
\newblock In {\em Planck Scale Effects in Astrophysics and Cosmology}, volume
  669 of {\em Lecture Notes in Physics}, pages 131--159. Springer, Berlin,
  Heidelberg.

\bibitem[Magueijo and Smolin, 2002]{MS}
Magueijo, J. and Smolin, L. (2002).
\newblock Lorentz invariance with an invariant energy scale.
\newblock {\em Physical Review Letters}, 88:190403.

\bibitem[Marino, 2025]{Marino2025}
Marino, F. (2025).
\newblock Phonons mimicking doubly special relativity kinematics.
\newblock {\em Physics Letters B}, 866:139529.

\bibitem[Moshinsky and Szczepaniak, 1989]{Moshinsky1989}
Moshinsky, M. and Szczepaniak, A. (1989).
\newblock The dirac oscillator.
\newblock {\em Journal of Physics A: Mathematical and General},
  22(17):L817--L819.

\bibitem[Mostafazadeh, 2002]{Mostafazadeh2002}
Mostafazadeh, A. (2002).
\newblock Pseudo-hermiticity versus {PT}-symmetry: The necessary condition for
  the reality of the spectrum of a non-hermitian hamiltonian.
\newblock {\em Journal of Mathematical Physics}, 43(1):205--214.

\bibitem[Pauli and Weisskopf, 1934]{PW34}
Pauli, W. and Weisskopf, V.~F. (1934).
\newblock {\"U}ber die quantisierung der skalaren relativistischen
  wellengleichung.
\newblock {\em Helvetica Physica Acta}, 7:709--731.
\newblock No DOI available (historic journal issue).

\bibitem[Relancio, 2022]{Relancio2022}
Relancio, J.~J. (2022).
\newblock Relativistic deformed kinematics: From flat to curved spacetimes.
\newblock {\em International Journal of Modern Physics D}, 31(11):2230004.

\bibitem[Relancio and Liberati, 2022]{RelancioLiberati2022}
Relancio, J.~J. and Liberati, S. (2022).
\newblock Black hole surface gravity in doubly special relativity geometries.
\newblock {\em Universe}, 8(2):136.

\bibitem[Tao et~al., 2022]{Tao2022Rainbow}
Tao, Y., Mu, B., Hui, S., and Tao, J. (2022).
\newblock Stationary and free-fall frame {K}err black hole in gravity's
  rainbow.

\end{thebibliography}

\end{document}